\PassOptionsToPackage{usenames,dvipsnames}{xcolor}
\documentclass[sigplan,screen, nonacm]{acmart}
\AtBeginDocument{%
  }

\usepackage{rcompcerto}
\usepackage{lstcoq}

\begin{document}

\title{RustCompCert: A Verified and Verifying Compiler for a Sequential Subset of Rust}


\author{Jinhua Wu$^*$, \hspace{0.2em} Yuting Wang$^*$, \hspace{0.2em} Liukun Yu$^*$, \hspace{0.2em} Linglong Meng$^{\mathsection}$}
\affiliation{
  \institution{Shanghai Jiao Tong University, China$^*$ \hspace{0.3em} University of Minnesota, USA$^{\mathsection}$}            
  \country{}                    
}
\email{yuting.wang@sjtu.edu.cn}            

\maketitle

\section{Introduction}

We aim to develop an end-to-end verified Rust compiler. It should
provide two guarantees: one is semantics preservation, i.e., the
behaviors of source code includes the behaviors of target code, with
which the properties verified at the source can be preserved down to
the target; the other is memory safety ensured by the verifying
compilation~\cite{tony-verifying}---the borrow checking
pass~\cite{borrow-check-doc}, which can simplify the verification of
Rust programs, e.g., by allowing the verification tools focus on the
functional correctness~\cite{Creusot, RustHorn, prusti, Aeneas, Verus,
  sound-borrow}.


However, there are significant challenges in verifying the Rust
compiler.  Some Rust compilation and checking passes, such as
\emph{Drop Elaboration}~\cite{elabdrop} and \emph{Borrow
  Checking}~\cite{borrow-check-doc}, involve complicated program
analysis algorithms. It is non-trivial to formalize and verify these
analyses within existing frameworks for verified compilers. Existing
formalizations of Rust~\cite{weiss2019oxide, rustbelt} primarily focus
on verifying its type soundness, rather than verifying the analysis
algorithms themselves. Aeneas~\cite{Aeneas, sound-borrow} proposes a
promising borrow checking algorithm based on a borrow-centric
semantics, but we focus on another direction---verifying the
algorithms used in the standard Rust compiler, namely
NLL~\cite{rust-nll} or Polonius~\cite{Polonius-update}.

%


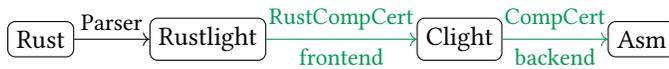
\begin{figure}[ht!]
\def\bheight{0.4cm}
\begin{tikzpicture}
    \hblock{\bheight}{draw, minimum width = 0.4cm, rounded corners=0.1cm}  (rust) {Rust};
    \hblock{\bheight}{draw, minimum width = 0.4cm, rounded corners=0.1cm, right = 1cm of rust}  (rustlight) {Rustlight};
    \hblock{\bheight}{draw, minimum width =0.4cm, rounded corners=0.1cm, right = 2cm of rustlight}  (clight) {Clight};
    \hblock{\bheight}{draw, minimum width = 0.4cm, rounded corners=0.1cm, right = 1.4cm of clight}  (asm) {Asm};

    \draw [->] (rust) -- node [above] {\small Parser} (rustlight);
    \draw [->, color=ForestGreen] (rustlight) -- node [above] {\small RustCompCert} node [below] {\small frontend} (clight);
    \draw [->, color=ForestGreen] (clight) -- node [above] {\small CompCert} node [below] {\small backend}  (asm);
 \end{tikzpicture}
\vspace{-0.4cm}
\caption{Structure of RustCompCert (\greentxt{green}: verified)}
\label{fig:rustcert-compiler}
\vspace{-0.1cm}
\end{figure}

We present our ongoing work on verifying a Rust compiler frontend in
Rocq and connecting it with CompCert~\cite{leroy09, Leroy-backend} to
build an end-to-end verified Rust compiler, which we call
RustCompCert. Its overall structure is illustrated
in~\figref{fig:rustcert-compiler}, which consists of three components:
(1) an unverified parser that translates source Rust code into
Rustlight, a sequential subset of Rust formalized in Rocq; (2) a
verified frontend that compiles Rustlight to Clight, including
Rust-specific passes such as Drop Elaboration and Borrow Checking; and
(3) a verified backend based on CompCert, an optimizing compiler that
compiles Clight to assembly\footnote{We use the framework of
  CompCertO~\cite{direct-refinement, compcerto}, an extension of
  CompCert, to support verified compositional compilation.}.

Verifying all features of Rust is not realistic. We therefore focus on
a core subset of the language, which we formalize as
Rustlight. Rustlight supports algebraic data types (i.e., struct and
enum), automatic memory management for Box via RAII and ownership
semantics, (im)mutable references with region (or lifetime)
annotations, as well as the composition of safe and unsafe modules
(e.g., safe Rust and C). The main unsupported features include
concurrency, polymorphism, traits, and higher-order functions (e.g.,
closures).


\paragraph{How we present it}
We plan to present our work in 20–25 minutes, with an additional 5–10
minutes reserved for Q\&A. The presentation will broadly follow the
structure of this paper. We will begin by introducing the compilation
passes of RustCompCert and its main correctness theorems
in~\secref{sec:rustcert}. We will then focus on how we formalize and
verify the borrow checking algorithm as described
in~\secref{sec:borrow-check}. Due to time and space limits, we may not
cover our approach to supporting unsafe code; interested readers can
refer to this paper~\cite{open-safety}. Our ongoing formalization is
publicly available at
\url{https://github.com/SJTU-PLV/CompCert/tree/rust-verified-compiler}.


%


\section{Overview of RustCompCert}\label{sec:rustcert}



\begin{figure}[ht!]
  \def\bheight{0.4cm}
\scalebox{0.9}{
\begin{tikzpicture}
    \hblock{\bheight}{draw, minimum width = 0.4cm, rounded corners=0.1cm}  (rl) {Rustlight};
    \hblock{\bheight}{draw, minimum width = 0.4cm, rounded corners=0.1cm, right = 1cm of rl}  (ir1) {RustIR};
    \hblock{\bheight}{draw, minimum width = 0.4cm, rounded corners=0.1cm, right = 1.6cm of ir1}  (ir2) {RustIR};
    \hblock{\bheight}{draw, minimum width = 0.4cm, rounded corners=0.1cm, right = 1.6cm of ir2}  (cl) {Clight};

    \draw [->] (rl) -- node [above] {\small Lower} (ir1);
    \draw [->] (ir1) -- node [above] {\small Drop} node [below] {\small Elaboration} (ir2);
    \draw [->] (ir2) -- node [above] {\small Clight} node [below] {\small Generation} (cl);

    \draw[->] (ir2.north west) to[bend left=130, looseness=1.5]
        node[midway, above] {\small Borrow Checking} (ir2.north east);
\end{tikzpicture}
}
\vspace{-0.5cm}
\caption{Structure of RustCompCert Frontend}
\label{fig:rustcert-frontend}
\vspace{-0.2cm}
\end{figure}
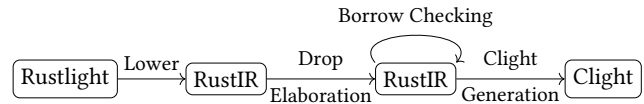

The internal structure of the RustCompCert frontend is shown
in~\figref{fig:rustcert-frontend}. We follow the design of the
standard Rust compiler (i.e., \code{rustc}) to structure our
compilation pipeline. The syntax of Rustlight is close to that of
surface Rust. We design an intermediate representation called
\emph{RustIR}, whose syntax is similar to MIR. The first pass lowers
Rustlight to RustIR in preparation for subsequent analyses,
corresponding to the MIR construction in
\code{rustc}~\cite{mir-construct}. We then perform Drop
Elaboration~\cite{elabdrop}, which is a key pass in \code{rustc} that
realizes the mechanism of automatic resource management. This pass
determines which \code{drop} operations (i.e., destructors) should be
executed by analyzing the initialization status of
variables. Consequently, the correctness of this pass crucially relies
on the soundness of this program analysis.

\vspace{-0.1cm}

We then perform borrow checking on RustIR. The borrow checker consists
of two components: a move checking pass~\cite{movecheck}, which
ensures that each variable is not used after being moved, and a borrow
checking pass based on Polonius~\cite{Polonius-update}. Compare to
NLL~\cite{rust-nll}, the stable borrow checker in \code{rustc} which
interprets regions as ``sets of program points that a loan may live'',
Polonius can be seen as an alias analysis, which is more suitable to
implement in the program analysis framework provided by CompCert. Its
interpretation of regions as ``sets of places that a reference may
point to'' aligns naturally with the values that may be stored in
references in the program semantics.
The Clight Generation pass serves as the code generation phase. For
example, it translates enum types into tagged unions and generates
drop glue functions for types that require destructors.

\paragraph{Compiler Correctness}
Like CompCert providing end-to-end semantics preservation for Clight
programs, RustCompCert also provides this guarantee for Rustlight
programs:
\begin{flalign*}
& \scalebox{0.9}{ $\forall\app (M_s: \textnormal{Rustlight})\app (M_t:
  \textnormal{Asm}),$} &&\\
 &\quad \scalebox{0.9}{$\code{RustCompCert}(M_s) = M_t \imply
  \csims{\sem{M_t}}{\sem{M_s}}$} &&
\end{flalign*}
 Here $\csims{\sem{M_t}}{\sem{M_s}}$ is defined as a standard backward
 simulation, meaning that the behaviors of source program is included
 in the behaviors of target program.


\paragraph{Soundness of Borrow Checking}
The memory safety ensured by the borrow checking has been utilized by
many Rust verification tools~\cite{Creusot, RustHorn, prusti, Aeneas,
  Verus, sound-borrow} to simplify the verification at the source
level, e.g., by focusing on its functional correctness. To provide
such guarantee, we extract a functional specification called
\emph{RustIRspec} which abstracts the low-level CompCert memory of the
RustIR semantics into a structured memory model. The soundness of
borrow checking is then formalized as a refinement between RustIRspec
and RustIR:
\begin{flalign*}
& \scalebox{0.9}{$
   \forall\app (M: \textnormal{RustIR}),\app
   \code{BorrowCheck}(M)\app \checkmark \imply$} &&\\
& \quad \scalebox{0.9}{$
  \csims{\seml{M}{\textnormal{RustIR}}}{\seml{M}{\textnormal{RustIRspec}}} \wedge \safe{\seml{M}{\textnormal{RustIRspec}}}$} &&
\end{flalign*}
For programs passing the borrow checking, we can conclude that its
specification is safe, i.e., it has no undefined
behaviors.~\footnote{In fact, we cannot show that the borrow checker
  rules out all undefined behaviors. For example, division-by-zero is
  a typical UB that borrow checking cannot prevent. Therefore, the
  safety ensured by the borrow checker is partial safety: some UBs may
  still occur in the semantics. To obtain total safety (absence of all
  UBs), it remains the prover’s responsibility to rule out these UBs.}
The predicate \code{safe} can be seen as a definition of partial
correctness~\cite{open-safety}. When the programs are open, it would
specify pre- and post-conditions to support composition.

\paragraph{End-to-end Verification}
By combining the soundness of borrow checking and the compiler
correctness, the properties verified in RustIRspec can be preserved to
the assembly:
\begin{align*}
 & \scalebox{0.9}{$
   \forall\app (M_s: \textnormal{RustIR})\app (M_t:
  \textnormal{Asm}),$} \\
& \quad \scalebox{0.9}{
  $\code{RustCompCert}(M_s) = M_t \imply    
  \safe{\seml{M_s}{\textnormal{RustIRspec}}} \imply \safe{\seml{M_t}{\textnormal{Asm}}}$}
\end{align*}
%

\section{Formalization and Verification of the Borrow Checking Algorithm}\label{sec:borrow-check}

\paragraph{Formalization of Polonius}
One possible strategy is to re-implement the current Polonius in Rocq,
for instance by porting its code from rustc. However, because it
relies on complex graph construction and solving
algorithms~\cite{rustc-polonius}, this approach would be difficult to
verify in Rocq in practice. We instead implement it within the Kildall
framework~\cite{kildall} in CompCert by defining its abstract domain
and transfer functions. The result of Polonius is then obtained as a
fixpoint computed by the Kildall solver.

\emph{Abstract Domain.}
The naive definition is a loans map from regions to set of loans (the
terminology representing borrowed places). However, borrow checking
relies on variance rules~\cite{rust-subtyping} to maintain equality
relationships between regions. The analysis must therefore not only
propagate loans across regions and program points, but also propagate
the equality relation between regions. If two regions are recorded as
equal, they must contain the same set of loans. We achieve this by
maintaining region equality in the abstract domain using a union–find
structure. To access the loans associated with a region, we use its
representative in the union–find to index the loans map instead of
using the region itself.
As a result, the abstract domain is a pair of type $(M \times U)$
consisting of a mapping from regions to loans and a union–find
structure. To integrate this domain into the Kildall framework, we
further prove that it forms a semi-lattice.


\emph{Transfer Function.}
A transfer function $F: \mathbb{N} \to (M \times U) \to (M \times U)$
describes how the abstract state of type $(M \times U)$ evolves for an
instruction at node $n \in \mathbb{N}$. It mainly specifies how loans
flow from one region (e.g., the assigner) to another (e.g., the
assignee). Moreover, new equality relations between regions may be
established (e.g., by variance
rules~\cite{rust-subtyping}). Conversely, some relations may need to
be removed when one of the regions in the relation becomes dead (i.e.,
will no longer be used). We design the transfer function following the
subtyping and reborrow rules in the NLL RFC~\cite{rust-nll} and their
implementation in \code{rustc}.

\vspace{-0.3cm}
\paragraph{Verification of Polonius}
Recall that we define the soundness of borrow checking as a refinement
between RustIRspec and RustIR. To prove this refinement for any
program that passes borrow checking, the key steps are: (1) to
establish an approximation relation between the abstract state
produced by the borrow-checking analysis and the concrete state (i.e.,
the memory model) in the semantics, and (2) to show that under this
relation every operation in the structured memory model of RustIRspec
can correspond to a non-stuck operation in the CompCert memory model
of RustIR. In particular, this implies the absence of memory errors.
The first step is the more challenging one. With Polonius viewed as a
form of alias analysis, we can relate each region to the set of memory
locations that may alias with the places containing this region, a
property that is much easier to formalize~\cite{verified-alias}. This
abstraction relation also guides potential improvements to the
Polonius algorithm in terms of precision and performance. For
instance, we can relax the alias sets to reduce precision in exchange
for better performance (e.g., by making Polonius field-insensitive),
while still being able to prove soundness using the same proof
technique.



\bibliography{refs}

@string{SPRINGER="Springer-Verlag"}

@string{ACM="Association for Computing Machinery"}

@string{CACM="Communications of the {ACM}"}

@string{NY="NY"}

@string{POPL=" ACM Symposium on Principles of Programming Languages"}

@string{PLDI21="Proc. 2021" # PLDI # " (PLDI'21)"}

@string{ICFP=" ACM SIGPLAN International Conference on Functional Programming"}

@article{leroy09,
  author = {Xavier Leroy},
  title = {Formal Verification of a Realistic Compiler},
  doi = {https://doi.org/10.1145/1538788.1538814},
  journal = CACM,
  year = 2009,
  volume = 52,
  number = 7,
  pages = {107--115}
}

@article{Leroy-backend,
  author = {Xavier Leroy},
  title = {A Formally Verified Compiler Back-end},
  journal = {Journal of Automated Reasoning},
  volume = 43,
  number = 4,
  pages = {363--446},
  year = 2009,
  doi = {https://doi.org/10.1007/s10817-009-9155-4},
  urllocal = {http://gallium.inria.fr/~xleroy/publi/compcert-backend.pdf},
  urlpublisher = {http://dx.doi.org/10.1007/s10817-009-9155-4},
  xtopic = {compcert}
}

@inproceedings{compcerto,
author = {Koenig, J\'{e}r\'{e}mie and Shao, Zhong},
title = {CompCertO: Compiling Certified Open C Components},
year = {2021},
isbn = {9781450383912},
publisher = {{ACM}},
address = {New York, NY, USA},
url = {https://doi.org/10.1145/3453483.3454097},
doi = {10.1145/3453483.3454097},
booktitle = PLDI21,
pages = {1095–1109},
numpages = {15}
}

@article{direct-refinement,
author = {Zhang, Ling and Wang, Yuting and Wu, Jinhua and Koenig, J\'{e}r\'{e}mie and Shao, Zhong},
title = {Fully Composable and Adequate Verified Compilation with Direct Refinements between Open Modules},
year = {2024},
issue_date = {January 2024},
publisher = {Association for Computing Machinery},
address = {New York, NY, USA},
volume = {8},
number = {POPL},
url = {https://doi.org/10.1145/3632914},
doi = {10.1145/3632914},
journal = {Proc. ACM Program. Lang.},
month = jan,
articleno = {72},
numpages = {31}
}

@article{RustBelt,
author = {Jung, Ralf and Jourdan, Jacques-Henri and Krebbers, Robbert and Dreyer, Derek},
title = {RustBelt: securing the foundations of the Rust programming language},
year = {2017},
issue_date = {January 2018},
publisher = {Association for Computing Machinery},
address = {New York, NY, USA},
volume = {2},
number = {POPL},
url = {https://doi.org/10.1145/3158154},
doi = {10.1145/3158154},
journal = {Proc. ACM Program. Lang.},
month = dec,
articleno = {66},
numpages = {34}
}

@misc{elabdrop,
  author = {{Rust Compiler Team}},
  howpublished = {},
  title = {Drop elaboration},
  url = {https://rustc-dev-guide.rust-lang.org/mir/drop-elaboration.html},
  year = {2025}
}

@misc{movecheck,
  author = {{Rust Compiler Team}},
  howpublished = {},
  title = {Tracking moves and initialization},
  url = {https://rustc-dev-guide.rust-lang.org/borrow_check/moves_and_initialization.html},
  year = {2025}
}

@misc{mir-construct,
  author = {{Rust Compiler Team}},
  howpublished = {},
  title = {MIR construction},
  url = {https://rustc-dev-guide.rust-lang.org/mir/construction.html},
  year = {2025}
}

@article{weiss2019oxide,
  title={Oxide: The essence of rust},
  author={Weiss, Aaron and Gierczak, Olek and Patterson, Daniel and Ahmed, Amal},
  journal={arXiv preprint arXiv:1903.00982},
  year={2019}
}

@article{sound-borrow,
author = {Ho, Son and Fromherz, Aymeric and Protzenko, Jonathan},
title = {Sound Borrow-Checking for Rust via Symbolic Semantics},
year = {2024},
issue_date = {August 2024},
publisher = {Association for Computing Machinery},
address = {New York, NY, USA},
volume = {8},
number = {ICFP},
url = {https://doi.org/10.1145/3674640},
doi = {10.1145/3674640},
journal = {Proc. ACM Program. Lang.},
month = aug,
articleno = {251},
numpages = {29}
}

@article{tony-verifying,
author = {Hoare, Tony},
title = {The verifying compiler: A grand challenge for computing research},
year = {2003},
issue_date = {January 2003},
publisher = {Association for Computing Machinery},
address = {New York, NY, USA},
volume = {50},
number = {1},
issn = {0004-5411},
url = {https://doi.org/10.1145/602382.602403},
doi = {10.1145/602382.602403},
journal = {J. ACM},
month = jan,
pages = {63–69},
numpages = {7}
}

@article{Aeneas,
author = {Ho, Son and Protzenko, Jonathan},
title = {Aeneas: Rust verification by functional translation},
year = {2022},
issue_date = {August 2022},
publisher = {Association for Computing Machinery},
address = {New York, NY, USA},
volume = {6},
number = {ICFP},
url = {https://doi.org/10.1145/3547647},
doi = {10.1145/3547647},
journal = {Proc. ACM Program. Lang.},
month = aug,
articleno = {116},
numpages = {31}
}

@article{Verus,
author = {Lattuada, Andrea and Hance, Travis and Cho, Chanhee and Brun, Matthias and Subasinghe, Isitha and Zhou, Yi and Howell, Jon and Parno, Bryan and Hawblitzel, Chris},
title = {Verus: Verifying Rust Programs using Linear Ghost Types},
year = {2023},
issue_date = {April 2023},
publisher = {Association for Computing Machinery},
address = {New York, NY, USA},
volume = {7},
number = {OPSLA1},
url = {https://doi.org/10.1145/3586037},
doi = {10.1145/3586037},
journal = {Proc. ACM Program. Lang.},
month = apr,
articleno = {85},
numpages = {30}
}

@inproceedings{Creusot,
author = {Denis, Xavier and Jourdan, Jacques-Henri and March\'{e}, Claude},
title = {Creusot: A Foundry for the Deductive Verification of Rust Programs},
year = {2022},
isbn = {978-3-031-17243-4},
publisher = {Springer-Verlag},
address = {Berlin, Heidelberg},
url = {https://doi.org/10.1007/978-3-031-17244-1_6},
doi = {10.1007/978-3-031-17244-1_6},
booktitle = {Formal Methods and Software Engineering: 23rd International Conference on Formal Engineering Methods, ICFEM 2022, Madrid, Spain, October 24–27, 2022, Proceedings},
pages = {90–105},
numpages = {16},
location = {Madrid, Spain}
}

@article{RustHorn,
author = {Matsushita, Yusuke and Tsukada, Takeshi and Kobayashi, Naoki},
title = {RustHorn: CHC-based Verification for Rust Programs},
year = {2021},
issue_date = {December 2021},
publisher = {Association for Computing Machinery},
address = {New York, NY, USA},
volume = {43},
number = {4},
issn = {0164-0925},
url = {https://doi.org/10.1145/3462205},
doi = {10.1145/3462205},
journal = {ACM Trans. Program. Lang. Syst.},
month = oct,
articleno = {15},
numpages = {54}
}

@misc{rust,
  author = {{The Rust Team}},
  howpublished = {},
  title = {The Rust Programming Language},
  url = {https://rust-lang.org/},
  year = {2025}
}

@misc{borrow-check-doc,
  author       = {The Rust Team},
  book         = {Rust Compiler Development Guide},
  title        = {MIR borrow check},
  url         = {https://rustc-dev-guide.rust-lang.org/borrow_check.html},
  year         = {2025}
}

@Misc{rustc-polonius,
  author = {{Rust contributors}},
  year = {2026},
  title = {Implementation of Polonius in Rust compiler},
  url = {https://github.com/rust-lang/rust/tree/main/compiler/rustc_borrowck/src/polonius},
  note = {Accessed February 2, 2026}
}

@Misc{Polonius-update,
  author = {{Rémy Rakic and Niko Matsakis on behalf of The Polonius Working Group}},
  year = {2023},
  title = {Polonius Update},
  url = {https://blog.rust-lang.org/inside-rust/2023/10/06/polonius-update.html},
  note = {Accessed Ocotober 12, 2025}
}

@Misc{rust-nll,
  author = {{Aaron Turon, Konrad Borowski, Hidehito Yabuuchi, and Dan Aloni}},
  year = {2017},
  title = {Non-Lexical Lifetimes},
  url = {https://rust-lang.github.io/rfcs/2094-nll.html},
  note = {Accessed Ocotober 12, 2025}
}

@misc{rust-subtyping,
  author = {{The Rust team}},
  howpublished = {},
  title = {Subtyping and Variance},
  booktitle = {The Rustonomicon},
  url = {https://doc.rust-lang.org/nomicon/subtyping.html},
  year = {2025}
}

@inproceedings{kildall,
author = {Kildall, Gary A.},
title = {A unified approach to global program optimization},
year = {1973},
isbn = {9781450373494},
publisher = {Association for Computing Machinery},
address = {New York, NY, USA},
url = {https://doi.org/10.1145/512927.512945},
doi = {10.1145/512927.512945},
booktitle = {Proceedings of the 1st Annual ACM SIGACT-SIGPLAN Symposium on Principles of Programming Languages},
pages = {194–206},
numpages = {13},
location = {Boston, Massachusetts},
series = {POPL '73}
}

@article{prusti,
author = {Astrauskas, Vytautas and M\"{u}ller, Peter and Poli, Federico and Summers, Alexander J.},
title = {Leveraging rust types for modular specification and verification},
year = {2019},
issue_date = {October 2019},
publisher = {Association for Computing Machinery},
address = {New York, NY, USA},
volume = {3},
number = {OOPSLA},
url = {https://doi.org/10.1145/3360573},
doi = {10.1145/3360573},
journal = {Proc. ACM Program. Lang.},
month = oct,
articleno = {147},
numpages = {30}
}

@misc{open-safety,
      title={End-to-end Compositional Verification of Program Safety through Verified and Verifying Compilation}, 
      author={Jinhua Wu and Yuting Wang and Liukun Yu and Linglong Meng},
      year={2025},
      eprint={2510.10015},
      archivePrefix={arXiv},
      primaryClass={cs.PL},
      url={https://arxiv.org/abs/2510.10015}, 
}

@InProceedings{verified-alias,
author="Robert, Valentin
and Leroy, Xavier",
editor="Hawblitzel, Chris
and Miller, Dale",
title="A Formally-Verified Alias Analysis",
booktitle="Certified Programs and Proofs",
year="2012",
publisher="Springer Berlin Heidelberg",
address="Berlin, Heidelberg",
pages="11--26",
isbn="978-3-642-35308-6"
}

\end{document}